\documentclass[aps,prb,superscriptaddress,floatfix,twocolumn]{revtex4}

\usepackage[dvips]{graphicx}
\usepackage{epsfig}
\usepackage{amsmath,amssymb}
\usepackage{bm}

\newcounter{Figure}
\newenvironment{FigureCaptions}{\begin{list}{
  Fig. \theFigure. \rm}
  {\protect\usecounter{Figure}\setlength{\labelwidth}{9em}}
  }{\end{list}}

\newcommand{\br}{\bm{r}}
\newcommand{\bu}{\bm{u}}
\newcommand{\D}{\text{d}}
\newcommand{\lrho}{\tilde{\rho}}

\newcommand{\fig}[2]
{
\noindent
Fig. #2
\\
\\
\\
\includegraphics[width=8.6cm]{#1}

\vfill

\clearpage
}

\begin{document}

\title{Theory of Liquid Crystal Anchoring at a Porous Surface}

\author{David L. Cheung}
\email{david.cheung@warwick.ac.uk}
\affiliation{Department of Physics and Centre for Scientific Computing,
  University of Warwick, Coventry, CV4 7Al, UK}

\begin{abstract}

  Using classical density functional theory (DFT) the effect of bringing a
  liquid crystal (LC) into contact with a porous substrate or matrix is
  investigated. The DFT used is a combination of the Onsager
  approximation to evaluate the excess free energy of the LC fluid and
  quenched annealed DFT to evaluate the interaction between the fluid
  and the porous substrate. When the fluid alignment far from the substrate is
  held perpendicular to its surface there is a thin layer of fluid
  aligned parallel to the substrate surface for low matrix
  densities. For higher matrix densities this director deformation
  propagates into the bulk of the fluid. Consideration of a system without confining walls suggests that for low matrix densities normal alignment is metastable compared to parallel alignment, while for higher matrix densities it is unstable. 

\end{abstract}
\maketitle

\section{Introduction}

The interaction between liquid crystals (LC) and solid substrates of
of great interest, both scientific and technological \cite{jerome1992a}.
The presence of the surface both breaks the symmetry of the LC phases
and often leads to alignment in a given direction. This tendency, often
called anchoring, is vital to the use of LC in display applications and
can be profoundly influenced by the structure of the surface.

Commonly surfaces may be rough or porous, e.g. Si0  \cite{barberi1990a,martinot-lagarde1996a}, which can lead to deviations
from the behaviour expected from smooth surfaces
\cite{sluckin1995a}. A porous substrate or matrix may be regarded as a
system of immobilised particles. Recent 
simulations of LC near rough walls \cite{cheung2005a,cheung2006a}
using such a model have shown 
that properties such as the anchoring coefficient and transition pressures
of a confined LC are influenced by the roughness of the
substrate. Due to the computational expense only a few surfaces at a
given roughness were studied. It would thus be desirable to study a larger 
(ideally infinite) number of surfaces.

One route to this is through replica or quenched annealed density
functional theory (QA-DFT) 
\cite{schmidt2002a,schmidt2003a,schmidt2004a,schmidt2005a}. In this the system 
comprises two components: the first, the quenched component, models the porous 
substrate, the second, the annealed component, models the fluid. The density 
distribution of the matrix, averaged over all matrix realisations, is 
determined though minimisation of a grand potential in the absence of the 
fluid. The density distribution of the fluid is then found by minimising a 
grand potential containing both matrix and fluid densities. This theory is fully outlined in the following section, along with details of the systems studied. The results that ensue from applying it to LC anchoring at porous surfaces are outlined in Sec.~\ref{sec:results}. Finally brief summary of this work and suggestions for future work are given in Sec.~\ref{sec:conclusions}

\section{Theory and model}
\label{sec:theory}

\subsection{Density functional theory}

In QA-DFT the grand potential is the sum of the grand potential of the
matrix alone and the grand potential of the fluid and matrix together,
i.e.
$\Omega[\rho_{\text{m}}(\br,\bu),\rho_{\text{f}}(\br,\bu)]=\Omega_{\text{m}}[\rho_{\text{m}}(\br,\bu)]+\Omega_{\text{f}}[\rho_{\text{m}}(\br,\bu),\rho_{\text{f}}(\br,\bu)]$
\cite{schmidt2002a}. Here we are interested in uniaxial molecules
characterised by a position $\br$ and orientation $\bu$. For an ideal matrix (i.e. with vanishing interactions between the matrix particles) $\Omega_{\text{m}}[\rho_{m}(\br,\bu)]$ is 
\begin{widetext}
\begin{equation}
  \beta\Omega_{\text{m}}[\rho_{\text{m}}(\br,\bu)]=
  \int \D\br \D\bu \rho_{\text{m}}(\br,\bu)\left\{
    \log\rho_{\text{m}}(\br,\bu)-1+\beta V_{\text{m}}(\br,\bu)-\beta\mu_{\text{m}}\right\}
\end{equation}
\end{widetext}
where $V_{\text{m}}(\br,\bu)$ is the external potential acting on the matrix, $\mu_{\text{m}}$ is its chemical potential, and $\beta=1/k_{\text{B}}T$. If the external potential
$V_{\text{m}}(\br,\bu)$ is used to confined the matrix to a region of space
then the matrix grand potential may be minimised analytically to give
$\rho_{\text{m}}(\br,\bu)=\rho_{\text{m}}=\exp(\beta\mu_{\text{m}})$ inside this region and 0 outside it. 

The grand potential for the fluid component is
\begin{widetext}
\begin{eqnarray}
  \beta\Omega_{\text{f}}\left[\rho_{\text{f}}(\br,\bu),\rho_{\text{m}}(\br,\bu)\right]&=&
    \int\;\D\br \D\bu \rho_{\text{f}}(\br,\bu)\left[
    \log\rho_{\text{f}}(\br,\bu)-1\right]+
  \beta\int\;\D\br \D\bu \left[V(\br,\bu)-\mu\right]
  \rho_{\text{f}}(\br,\bu) \nonumber\\
  &+&\beta F_{\text{ex}}\left[\rho_\text{f}(\br,\bu)\right]+
  \beta
  F_\text{ex}^{mf}\left[\rho_\text{f}(\br,\bu),\rho_{\text{m}}(\br,\bu)\right]
  \label{eqn:gp_f}
\end{eqnarray}
\end{widetext}
where $\rho_\text{f}(\br,\bu)$ is the density distribution of the fluid
component, $V(\br,\bu)$ is the external potential acting upon it,
and $\mu$ is the fluid chemical potential.
$F_\text{ex}\left[\rho_\text{f}(\br,\bu)\right]$ and
$F_\text{ex}^{mf}\left[\rho_\text{f}(\br,\bu),\rho_\text{m}(\br,\bu)\right]$ are the
excess free energies due to interactions between the fluid molecules
and fluid-matrix molecules respectively. From hereon in the subscript
f denoting the fluid component will be omitted. The excess FE is
evaluated within the Onsager approximation\cite{onsager1949a}
\begin{eqnarray}
  \beta F_\text{ex}\left[\rho(\br,\bu)\right]&=&
  -\frac{1}{2}\int \D\br_1\D\bu_1\D\br_2\D\bu_2 \nonumber\\
  & &
  f(\br_{12},\bu_1,\bu_2)\rho(\br_1,\bu_1)\rho(\br_2,\bu_2)
\end{eqnarray}
where $f(\br_{12},\bu_1,\bu_2)$ is the Mayer function. For convenience we assume that the interaction between the matrix and fluid particles is identical to the interactions between the fluid particles. 

In this work we are only concerned with systems that vary in
the $z$ direction only; the surfaces are assumed to be homogeneous in
the $x$ and $y$ directions. The method for finding the equilibrium
density is the same as in previous work \cite{allen2002a,cheung2004a}; the density and its logarithm are
expanded in spherical harmonics 
\begin{subequations}
  \label{eqn:expansions}
  \begin{align}
  \rho(z,\bu)=\sum_{\ell\text{m}} \rho_{\ell\text{m}}(z)Y_{\ell m}^*(\bu) \\
  \log\rho(z,\bu)=\sum_{\ell\text{m}}\lrho_{\ell m}(z)Y_{\ell\text{m}}(\bu)
  \end{align}
\end{subequations}
where $Y_{\ell\text{m}}(\bu)$ is a spherical harmonic.  The Mayer function is
expanded in rotational invariants. Inserting these into the grand
potential (Eq.~\ref{eqn:gp_f}) and integrating over angles and the $x$ and $y$ axes gives 
\begin{widetext}
\begin{eqnarray}
  \beta\Omega_\text{f}[\rho(\br,\bu),\rho_\text{m}(\br,\bu)]&=&
  \int\;\D z\;\sum_{\ell\text{m}} \rho_{\ell\text{m}}(z)\left\{\lrho_{\ell\text{m}}(z)+
  \beta V_{\ell\text{m}}(z)-\sqrt{4\pi}(1+\beta\mu)\delta_{\ell0}\right\} \nonumber\\
  &-&\int\;\D z_1\D z_2\; \sum_{\ell_1\ell_2\text{m}}A_{\ell_1\ell_2\text{m}}(z_1-z_2)\rho_{\ell_1\text{m}}(z_1)\left(\rho_{\ell_2\text{m}}(z_2)-\frac{1}{\sqrt{4\pi}}\rho_{\text{m}}(z_2)\delta_{\ell_20}\right) \nonumber\\
  & &
\end{eqnarray}
\end{widetext}
where $V_{\ell\text{m}}(z)$ are the spherical harmonics coefficients of the
external potential and $A_{\ell_1\ell_2\text{m}}(z_1-z_2)$ are the spherical
harmonics coefficients of the excluded area \cite{allen2002a}. $\rho_\text{m}(z)$ is the matrix
density, equal to $\rho_\text{m}$ for $z_\text{min}\le z\le z_\text{max}$ and 0 otherwise.
The equilibrium fluid density is then found by numerically minimising
this with respect to $\lrho_{\ell\text{m}}(z)$ using the conjugate gradients
method \cite{numericalrecipies}. When needed $\rho_{\ell\text{m}}(z)$
are found from Eqs.~\ref{eqn:expansions}. Once the
$\rho_{\ell\text{m}}(z)$ that minimises $\Omega_\text{f}$ has been
found the fluid density profile is given by
$\rho(z)=\int\;\D\bu\;\rho(z,\bu)$. The orientational ordering of the
fluid is described through the ordering tensor
\begin{equation}
  Q_{\alpha\beta}(z)=
  \tfrac{3}{2}\int
  \;\D\bu\;\rho(z,\bu)u_\alpha u_\beta-
  \tfrac{1}{2}\delta_{\alpha\beta},\qquad\alpha,\beta=x,y,z \;.
\end{equation}
The order parameter is given by the largest eigenvalue of $Q_{\alpha\beta}(z)$ and the director by the eigenvector associated with it.

Both the fluid and the matrix particles are modelled as hard
ellipsoids of revolution of elongation $e=a/b=15$. The chemical potential is set to $\mu=2.0k_B$ well inside the nematic phase. This class of model
have been well studied as model liquid crystals \cite{allen1993a}. Two
different systems were considered. In the first the fluid is confined
between two hard walls at $z=0$ and $z=L=200b$, with the matrix confined in
the region $0\le z \le z_\text{max}=30b$. The wall at $z=L$ is used to provided
(homeotropic or planar) alignment far from the matrix. In the second
system the matrix is placed in the centre of a periodic system of
width $200b$. The fluid density is then minimised from starting states parallel and normal to the matrix surface.

\section{Results}
\label{sec:results}

\subsection{Confined Geometry}

Shown in Fig.~\ref{fig:confined_density_and_order_profiles} are the density 
and order parameter profiles for the slab geometry with both planar and 
homeotropic alignment at the far wall. In all cases far from the matrix 
$\rho(z)$ is constant. For the lowest matrix density ($\rho_{\text{m}}=0.05b^{-3}$) 
the fluid density with the matrix is non-zero, although far lower than in 
the fluid outside the matrix. At larger $\rho_\text{m}$ the fluid density inside the 
matrix is zero. The variation in $\rho(z)$ near the matrix surface depends 
on $\rho_m$ and the alignment far from the matrix.  For planar alignment 
(Fig.~\ref{fig:confined_density_and_order_profiles}a) and $\rho_\text{m}<0.25b^{-3}$ 
$\rho(z)$ drops smoothly from the bulk value at $z\approx45b$ (one molecular 
length from the matrix surface) to 0 just inside the matrix. For larger values 
of $\rho_\text{m}$ a peak appears in $\rho(z)$ at $z\approx37.5b$ (half a molecular 
length from the surface), corresponding to an absorbed layer of molecules. $\rho(z)$ then drops sharply to 0 about 
one molecular diameter outside the matrix. For these $\rho_{\text{m}}$ the matrix becomes impenetrable to the fluid and starts to behave like a rough hard wall \cite{schmidt2003a}. When the far wall gives rise to 
homeotropic alignment $\rho(z)$ has a slight peak at $z\approx40b$ for all $\rho_\text{m}$ that grows stronger for increasing $\rho_\text{m}$. For the highest values of $\rho_\text{m}$ studied however, this peak is weaker than in the corresponding system with planar alignment.

Shown in Fig.~\ref{fig:confined_density_and_order_profiles}c are order 
parameter profiles ($S(z)$) for the confined LC with a far planar wall. 
In common with the density profiles these are constant far from the matrix, while 
for all $\rho_\text{m}$ $S(z)=0$ inside the matrix. For $\rho_\text{m}=0.05b^{-3}$ $S(z)$ 
drops smoothly to 0 while for a slightly higher $\rho_\text{m}$ a sharp peak 
appears just inside the matrix surface. At higher $\rho_\text{m}$ $S(z)$ also drops 
smoothly from its bulk value to 0, although this occurs outside the matrix 
and over a smaller range of $z$. $S(z)$ for the far homeotropic wall is 
shown in Fig.~\ref{fig:confined_density_and_order_profiles}d. The behaviour 
of $S(z)$ near the matrix surface differs markedly for different $\rho_\text{m}$.  
For low $\rho_\text{m}$ there are two peaks in $S(z)$ either side of the matrix 
surface, while deeper in the matrix $S(z)$ goes to zero. At larger $\rho_\text{m}$ 
the order parameter drops to 0 at the matrix surface.

For the system with far homeotropic alignment the behaviour of $S(z)$ near the 
matrix surface in influenced by the behaviour of the director. Plotted in 
Fig.~\ref{fig:confined_director} is $\theta(z)$, the angle between the 
director and the $z$ axis (the wall normal). For low $\rho_\text{m}$ the director 
lies parallel to the z-axis ($\theta(z)=0$) far from the matrix surface. Near the matrix 
surface however, it abruptly changes to lie in the $xy$ plane ($\theta(z)=90^\circ$). This preference 
for parallel alignment may be understood as the matrix consists of disordered 
ellipsoids, so the interface between it and the fluid is equivalent to an 
isotropic-nematic interface. Previous studies of this and similar models
 \cite{koch1999a,allen2000a} have shown that the director prefers to lie in 
the plane of the interface  For $\rho_\text{m}>0.26$ the preference for 
planar anchoring at the matrix surface leads to continuous director variation 
through the cell from homeotropic at the far wall to planar at the matrix 
surface. By contrast for the far planar wall is constant $\theta(z)=90^\circ$ for all 
$\rho_\text{m}$.

\subsection{Open Geometry}

Density and order parameters for the open system are shown in 
Fig.~\ref{fig:open_density_and_order_profiles}. As can be seen for both the 
homeotropic
and planar anchoring $\rho(z)$ and $S(z)$ are similar to those for the 
confined system: far from the matrix these are constant while close to the 
surface of the matrix these drop to smaller values. Also, with the exception 
of the $\rho_\text{m}=0.05b^{-3}$ system, $\rho(z)$ drops to 0 within the matrix.

The behaviour of the director is different in the open system. When the initial configuration for the minimisation has the director aligned parallel to the matrix surface the director remains in this state. For an initial normal alignment the final state also has normal alignment when $\rho_\text{m}\le0.25b^{-3}$, while for higher $\rho_\text{m}$ the final state has parallel alignment. This suggests that the normal alignment is metastable with respect to parallel alignment. In order to confirm this we calculate the surface free energy $\gamma$, given by the excess (over bulk) grand potential per unit area \cite{hansenmcdonald}
\begin{equation}
  \beta\gamma=
  \frac{\beta\Omega\left[\rho(\br,\bu)\right]-\beta\Omega_b\left[\rho_b(\bu)\right]}{2A}
\end{equation}
where $\Omega_b\left[\rho_b(1)\right]$ is the bulk grand potential. This is plotted in Fig.~\ref{fig:open_gp} for both initial parallel and normal alignments. As may be seen for $\rho_\text{m}\le0.25b^{-3}$ $\gamma$ is lower for parallel alignment, indicating that normal alignment is metastable. For higher $\rho_\text{m}$ normal alignment is unstable and $\gamma$ is identical for both initial conditions.

\section{Conclusion}
\label{sec:conclusions}

In this paper the effect of bringing a liquid crystalline fluid into contact with a porous substrate has been studied using QA-DFT. It was found that aligning the LC parallel to the substrate surface is favourable to normal alignment. For matrices of low density normal alignment is found to be metastable, while for higher $\rho_m$ it is unstable. This preference for parallel alignment is in agreement with studies of the nematic-isotropic interface for the same model.

There are a number of possible extensions to this work. It would be of
interest to compare the QA-DFT results to those of computer
simulations. Also interesting phenomenon in the vicinity of the
nematic-isotropic transition that may be examined or for LC mixtures and
for different interactions between the fluid and matrix particles.

\section*{Acknowledgements}

The author would like to thank Mike Allen and Matthias Schmidt for helpful advice. This work was funded by UK EPSRC and computational resources were provided by the Centre for Scientific Computing, University of Warwick.

\section*{Figure Captions}
\begin{FigureCaptions}
\item
\label{fig:confined_density_and_order_profiles}
(a) Density profiles ($\rho(z)$) for confined LC with far planar wall. 
Solid line $\rho_\text{m}=0.05b^{-3}$, dotted line $\rho_\text{m}=0.10b^{-3}$, dashed line 
$\rho_\text{m}=0.25b^{-3}$, long dashed line $\rho_\text{m}=0.30b^{-3}$, and dot-dashed line 
$\rho_\text{m}=0.40b^{-3}$.  
(b) Density profiles for confined LC with far homeotropic wall. Symbols as in 
(a).
(c) Order parameter profiles ($S(z)$) for confined LC with far planar wall. 
Symbols as in (a).
(d) Order parameter profiles for confined LC with far homeotropic wall. 
Symbols as in (a).

\item
\label{fig:confined_director}
Director angle $\theta(z)$ for confined LC with far homeotropic wall. Solid 
line shows $\rho_\text{m}=0.05b^{-3}$, dotted line shows $\rho_\text{m}=0.25b^{-3}$, dashed 
line shows $\rho_\text{m}=0.26b^{-3}$, long dashed line shows $\rho_\text{m}=0.30b^{-3}$ and 
dot-dashed line shows $\rho_\text{m}=0.40b^{-3}$.

\item
\label{fig:open_density_and_order_profiles}
(a) Density profiles for open system with parallel alignment. Solid line $\rho_\text{m}=0.05b^{-3}$, dotted line $\rho_\text{m}=0.10b^{-3}$, dashed line 
$\rho_\text{m}=0.25b^{-3}$, long dashed line $\rho_\text{m}=0.30b^{-3}$, and dot-dashed line 
$\rho_\text{m}=0.40b^{-3}$.
(b) Density profiles for open system with perpendicular alignment.
(c) Order parameter profiles for open system with parallel alignment. Symbols 
as in (a).
(d) Order parameter profiles for open system with perpendicular alignment.

\item
\label{fig:open_gp}
Surface free energy ($\gamma=\Omega-\Omega_b$) per unit length for open systems with parallel (solid line) and normal (dashed line).

\end{FigureCaptions}
\newpage
\fig{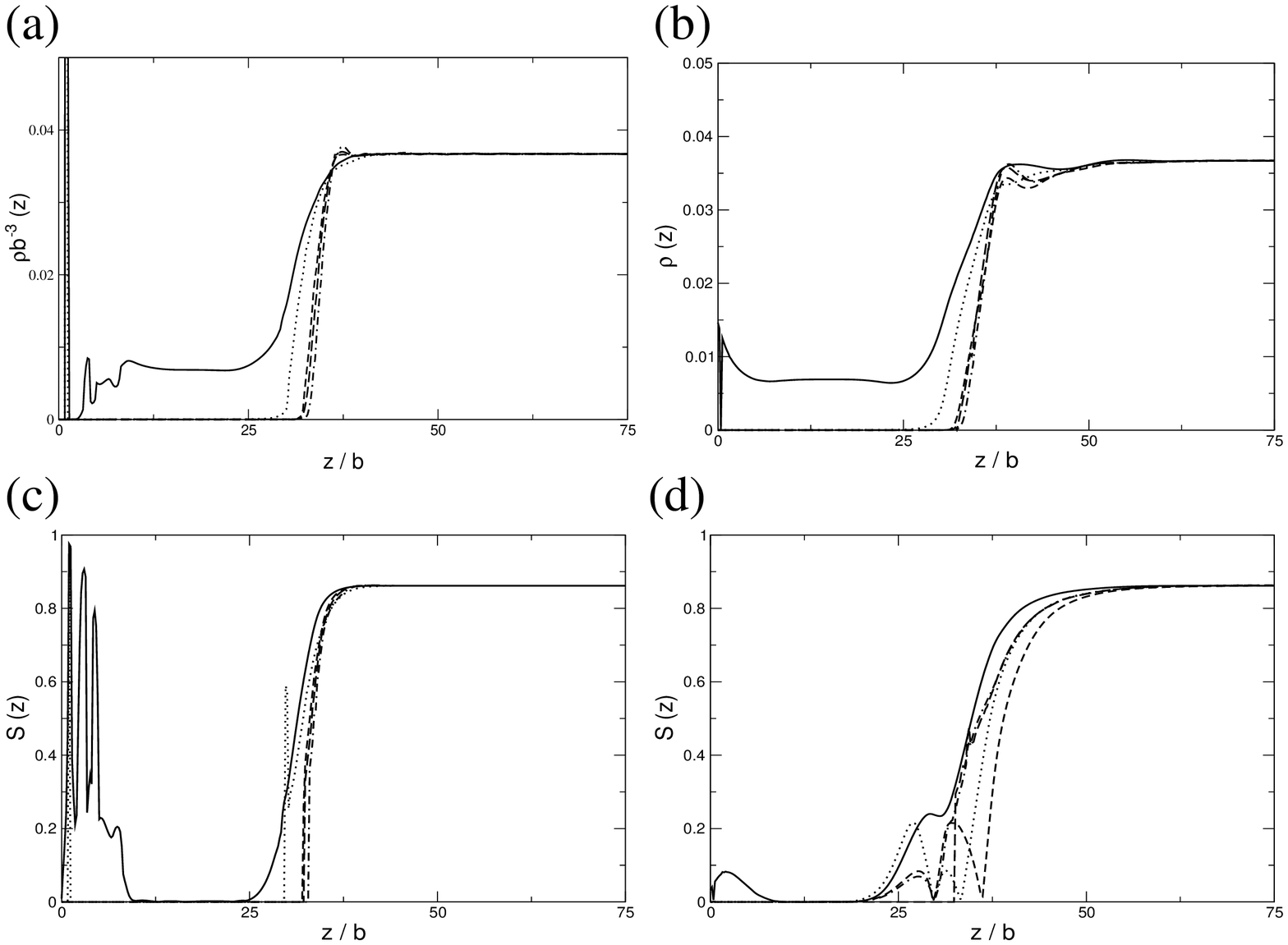}{1}
\fig{fig2.eps}{2}
\fig{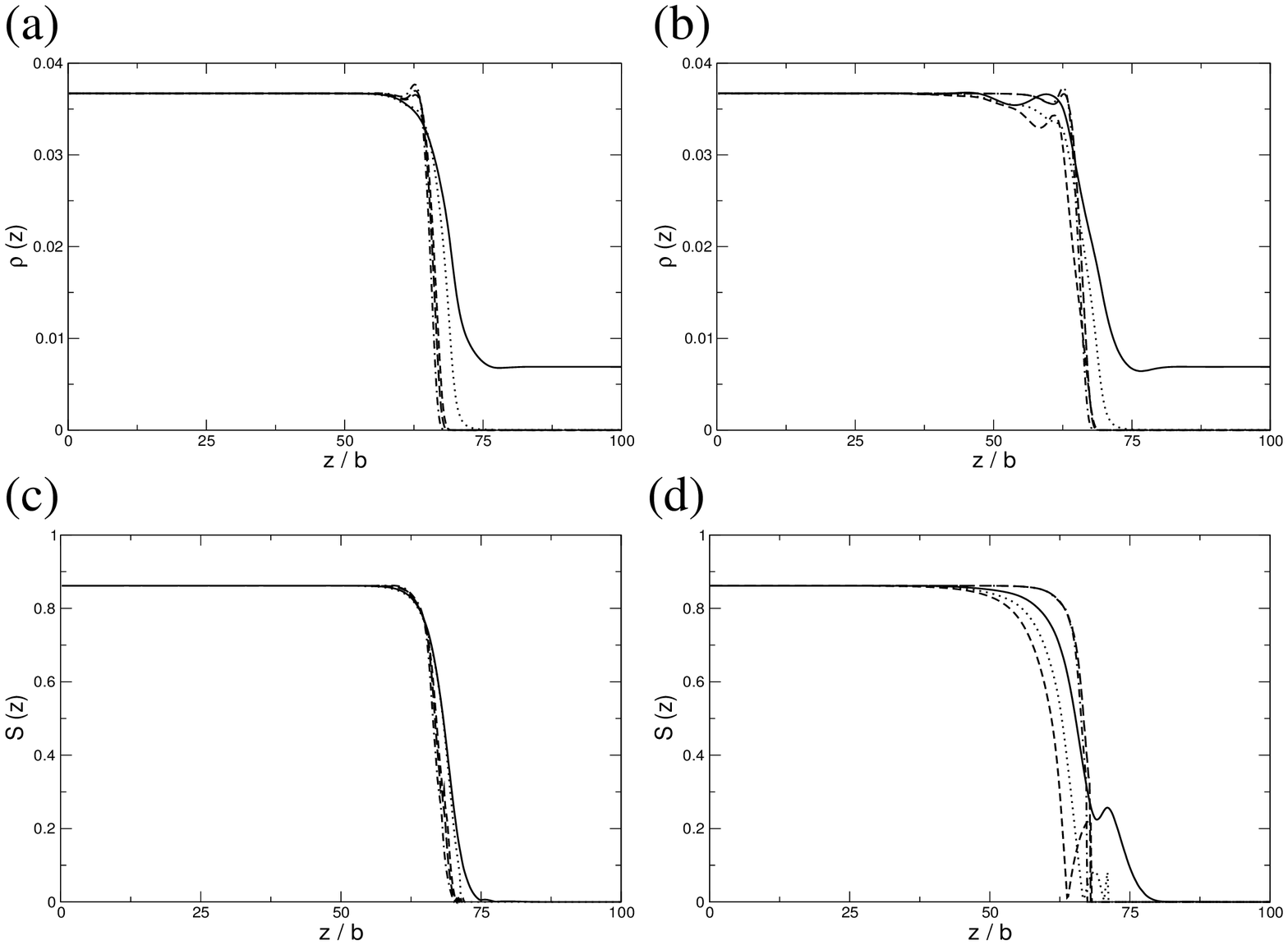}{3}
\fig{fig4.eps}{4}

\end{document}